# Discovering Drug-Drug and Drug-Disease Interactions Inducing Acute Kidney Injury Using Deep Rule Forests

Bowen Kuo
*Department of Information Management*
*National Sun Yat-sen University*
Kaohsiung, Taiwan
bowenkuo@outlook.com

Yihuang Kang
*Department of Information Management*
*National Sun Yat-sen University*
Kaohsiung, Taiwan
ykang@mis.nsysu.edu.tw

Pinghsung Wu
*Division of Nephrology*
*Kaohsiung Medical University*
Kaohsiung, Taiwan
970392@kmuh.org.tw

Sheng-Tai Huang
*Department of Information Management*
*National Sun Yat-sen University*
Kaohsiung, Taiwan
shengtai.huang28@gmail.com

Yajie Huang
*Department of Information Management*
*National Sun Yat-sen University*
Kaohsiung, Taiwan
judy89230@gmail.com

*Abstract*—Patients with Acute Kidney Injury (AKI) increase mortality, morbidity, and long-term adverse events. Therefore, early identification of AKI may improve renal function recovery, decrease comorbidities, and further improve patients' survival. To control certain risk factors and develop targeted prevention strategies are important to reduce the risk of AKI. Drug-drug interactions and drug-disease interactions are critical issues for AKI. Typical statistical approaches cannot handle the complexity of drug-drug and drug-disease interactions. In this paper, we propose a novel learning algorithm, Deep Rule Forests (DRF), which discovers rules from multilayer tree models as the combinations of drug usages and disease indications to help identify such interactions. We found that several disease and drug usages are considered having significant impact on the occurrence of AKI. Our experimental results also show that the DRF model performs comparatively better than typical tree-based and other state-of-the-art algorithms in terms of prediction accuracy and model interpretability.

## I. INTRODUCTION

Acute kidney injury (AKI) is a severe adverse event characterized by rapid decline of renal function, as a leading cause of complications, disability, and mortality [1]. Acute care, intensive care unit, or general ward often encounter patients with AKI, caused by underlying medical conditions or inappropriate drug usage. Despite there being standard care for patients with AKI, the mortality rate remains high. The mortality rate in severe AKI is about 50% to 60% [2]. Therefore, early diagnosis and treatment of AKI could help renal recovery, avoid complications, and improve patients' survival rate.

Nephrotoxic drugs are involved in 19-25% of the cases with severe AKI in critically ill patients [3], [4]. However, AKI caused by drug-drug interactions (DDI) or drug-disease interactions (DDX) is preventable, as many such interactions can be identified, and drugs can be substituted with less-nephrotoxic ones. The reversibility of AKI is related to the duration of risk factors exposure and the severity of the injury [5]. Therefore, predicting and avoiding unexpected AKI is substantial for clinical practice. DDI may occur in simultaneous drug exposure, which causes decreasing or increasing the action of multiple drugs [6]. For example, the drug combinations of angiotensin-converting enzyme inhibitor (ACEI)/Angiotensin II Receptor Blocker (ARB), diuretic, and nonsteroidal anti-inflammatory drug (NSAID) could precipitate AKI. This concurrent drug use is also known as "triple whammy" to the kidney [7]. As a consequence, healthcare providers should inform patients with kidney dysfunction to avoid NSAID exposure.

It is difficult for typical statistical methods, such as logistic regression models, to identify complex DDIs and DDXs. Patients with higher disease severity usually have more comorbidities and drug prescriptions, so many interactions between diseases and drugs could be found. We propose a novel approach, Deep Rule Forests, based on random forest [8] and a deep architecture-like ensemble learning algorithm to identify significant DDIs and DDXs. Our proposed approach with layer-wise rule forests can discover latent interactions/rules that potentially trigger AKI.

The rest of this paper is organized as follows. In Section II, we discuss and review background related to DDI and DDX identifications along with machine learning algorithms in literature. We elaborate our proposed approach in Section III. In Section IV, we present our findings with the evaluations of model performance and interpretability. We conclude and summarize in Section V.

## II. BACKGROUND AND RELATEDD WORK

There are more than 5,000 and 295 cases per million people per year requiring non-dialysis care and dialysis treatment in the U.S., respectively [9]. Prescribed drugs, such as radiocontrast agents, NSAIDs, antibiotics (β-lactam inhibitors, aminoglycosides, sulphonamides), antifungal agents (amphotericin), anti-viral agents (acyclovir), antihypertensive drugs (ACEI, ARB) are found contributing to AKI [10]. However, how to detect the AKI risk among the combinations of these drugs is still little known.

Inappropriate drug prescription is one of the most common causes of AKI. For example, the prescription of diuretics, ACEIs/ARBs, or NSAIDs in volume-depleted patients can exacerbate the AKI event [11]. These drugs can decrease renal perfusion and the amount of glomerular filtration rate. Besides, antibiotics, such as beta-lactam inhibitors, aminoglycosides, and sulfonamides, may cause acute interstitial nephritis. Furthermore, AKI caused by post-renal obstruction could be found in acyclovir (anti-viral agent) or Sulfonamide (antibiotics) in patients with inadequate fluid status [12].

Beyond the well-known nephrotoxic agents, several common prescription drugs were also found potentially nephrotoxic. For example, Dormuth et al. used a logistic regression model to estimate the increased risk of AKI in high potency statins (lipid-lowering drug) [13]. Besides, Shih et al. found the increased risk of AKI among dipeptidyl peptidase-4 inhibitor (glucose-lowering drug) users compared to nonusers in a conditional logistic regression model with confounders adjustment [14]. Furthermore, McCoy et al. employed univariate comparisons between the pre-intervention and post-intervention periods with the Pearson Chi-square test for categorical variables and the log-rank test for the time from events to predict response variables. These studies are precise to measure the modification or discontinuation rate per 100 events for medications included in the interruptive alert within 24 hours of increasing creatinine improved [15]. However, these studies did not provide specific information on co-occurrence drug exposure and the risk of AKI.

Drug-drug interactions (DDIs) occur when two or more drugs react with each other, which may cause increasing or decreasing drug reaction, unexpected side effects, and even death of the patients [16]. DDIs are the crucial issues in clinical practice, accounting for 2-6% of all hospital admissions, and estimating about 500 million pounds annual cost by UK National Health Service [17]. The use of interacting medications has increased since 2005, accounting for 15% of senior Americans at the risk increment of DDIs [18]. The more drugs the patients take, the more likely they would interact [19]. Juurlink et al. investigated the association between medication use and hospital admission due to drug toxicity. Most hospitalization of elderly patients occurs after the administration of a drug triggering drug-drug interaction. They concluded that many of these drug interactions could be avoided [20]. Yue et al. used the best-first strategy to search the space of feature subsets, and iteratively expand the subset according to the rule of best Merit value and continue to use lasso regression to select highly correlated features/variables. However, their results show that it can only increase the accuracy of the model prediction. We still do not know which combinations of drug usages might cause DDIs [21].

Besides, Adverse Drug Reaction (ADR) often results either from the prescription of a single drug or a particular class of DDIs [22]. Among them, taking nephrotoxic drugs is one of the high-risk factors for AKI. Nephrotoxic drugs involved in 19-25% of the cases with severe AKI in critically ill patients [3], [4]. Among patients with kidney diseases, polypharmacy is common to cause DDIs, which may cause unexpected adverse effects and even death [21]. Some ADRs are predictable and preventable, including known DDIs and DDXs [23]. For example, drug interaction between calcium channel blocker (blood pressure-lowering drug) and Clarithromycin (antibiotics) was found to increase the risk of AKI [24]. However, it is time-consuming to evaluate the DDIs and DDXs one by one because of the complexity of multiple comorbidities and polypharmacy in patients. Still, it is challenging for typical statistical approaches, such as generalized linear models) to take into account multiple interaction terms within one model. That is, pre-defining suitable interaction terms does help such statistical approaches get better prediction, but it is almost impossible and time-consuming for researchers to manually identify them.

Nevertheless, rule learning methods, such as RIPPER [25], rule-based C5.0 [26] and CART [27], may help automatically identify aforementioned interactions in patient prescription and healthcare history data. Such machine learning algorithms have been widely used in different fields, especially when interpretability is a concern. In rule-based learning algorithms, conjunctive clauses (AND-rules) and disjunctive clauses (OR-rules) are often one-level rules [28]. Among these rule learners, rules from tree-based methods are different from typical rule-based learning algorithms [28]. Tree-based methods mainly divide data into smaller subsets by choosing a series of split points known as rules. These rules are in conjunctive normal form (CNF; AND-of-ORs) and split data space into mutually exclusive regions as shown in Fig. 1. In Fig. 1, we can observe that the tree separates iris data [30] into three regions by choosing two split points. In other words, complicated interaction of feature status can be represented as regions, which provide more representational power than original features. This technique is established in the field of representation learning [31], the set of methods that identifies appropriate representations of data so that learning tasks can perform better in terms of accuracy than using the original data. Namely, the goal of representation learning is to automatically identify data transformations that map raw input features into representations that are more suitable for specific machine learning tasks. Therefore, in such cases, manual and cumbersome features extractions are not required [31]. In recent decades, ensemble learning with tree-based algorithms, such as random forest [8] and XGBoost [32], were proposed. These learning algorithms combine multiple trees into a forest with row resampling and column sampling techniques, and therefore they usually have more expressive power to discover signals/patterns in data. They are efficient rule learners, but their rules may conflict or overlap and thus the rules are hard to be directly interpreted by humans.

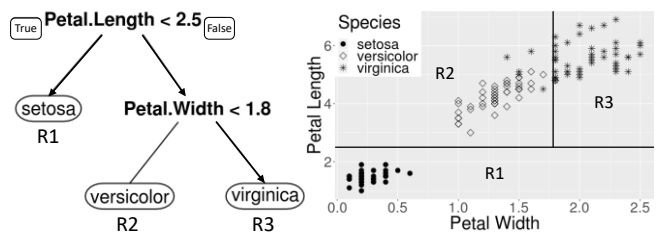

Fig. 1. The tree structure and the scatter plot of the decision tree applied to iris data [30]

Aside from obtaining diverse/distributed representations to increase expressive power, combining and reusing the existing representations is another way to boost the expressive power of a model. One popular example is the rise of deep neural networks and deep model architecture [33], [34], in which the network weights are used to perform linear and non-linear transformations and combinations in layer-wise fashions. Bengio et al. [35] argue that representation learning has two advantages: the first is that deep architecture promotes the reuse of features. The concept of reuse explains the power of distributed representation and is the core theory behind deep learning [34]. The second is that deep architecture leads to features that become progressively more abstract in higher-level representations. Observed empirically, layer-wise stacking of feature extractions sometimes produces better representations [31], [36]. On the other hand, researchers have proposed tree-based models with deep architectures, too. For example, Forward Thinking: Building Deep Random Forest (FTDRF) [37] and Deep Forest [38], have been proposed to stack multiple random forests into multilayer structures. In both algorithms, random forest in every layer will generate the predicted probability matrix as the new input data for the following layer. The probability matrices can be considered as a series of compressions of input features in the layer-wise fashion. These compressed representations can be further applied to other supervised machine learning methods, e.g. linear model, tree-based model, or multilayer perceptron (MLP) [34]. The aforementioned tree-based representation learning methods may potentially increase the accuracy of model prediction. However, they are still black-box models. The representations in the form of a probability matrix are incomprehensible, as they are simply the predicted probabilities of the target classes. Therefore, in order to provide interpretable and more expressive data representations, we propose our representation learning method that discovers DDIs and DDXs in disjunctive normal form (DNF; OR-of-ANDs) of rules. These discovered sets of rules are able to represent complex DDIs and DDXs. The importance of found DDIs and DDXs can also be further determined by other learning algorithms, such as linear models and tree-based models.

## III. DETECTION OF DDIS/DDXS FOR AKI—USING DEEP RULE FOREST

We propose a novel learning algorithm, Deep Rule Forest (DRF), that helps discover and extract rules from the combinations of disease conditions and drug usages. DRF maps data into rule representations using modified random forests in a layer-wise fashion. In Fig. 2, the table on the left represents the raw data, with each row indicating a doctor visiting record and each column indicating a patient feature (such as information on the patient's medical history and medications). After obtaining decision trees by applying random forest algorithm to the raw data, in the arrows (i), we map the observations of the training set into the containing regions with rules extracted from random forest. The arrow (ii) shows that the collated table of containing regions on the right side is prepared for generating another random forest in the next layer.

The granularity of rules in regions is determined by the number of tree splits. When a tree grows more deeply, it can handle more complex datasets but will also generate more complicated rules. The depth of decision trees controls how much information would be transmitted during rule mapping, which is also considered the model capacity. A fine-grained decision tree can map a dataset into relatively more regions, which raise the model capacity of information transmitting. However, if a decision tree grows too deep, it often tends to overfit the training data [39]. Therefore, the granularity of decision trees is a critical and considered a hyper-parameter of our proposed DRF. Fine-tuning hyper-parameters of the DRF can be considered as lowering validation errors step by step in deep neural networks [40]. Therefore, it depends on the difficulty of learning tasks.

We propose to construct the DRF model with multiple layers based on the notion of deep architecture [35]. The DRF automatically discovers rules from the variable-level features to the region-wise representations. Furthermore, hierarchical representation learning is constituted by the formation of higher-layer features through the compositions of low-layer features. Fig. 3 illustrates how to learn layer-wise distributed region representations of data. In the beginning of the training process, each tree in the 1st layer generates mutually exclusive regions/rules formed in CNF. Followed by the 1st layer, each tree in the successive layers performs splitting by choosing the appropriate set of regions in the selected trees in the previous layer. The procedure of finding appropriate sets of regions can reconstruct multiple CNF rules with OR-operators and reform the rules into DNF. In Fig. 3, take the first split of T2.1 as an example, DRF samples a set of trees from the 1st layer. Tree T1.1 and some branches/regions are selected as they can further minimize the Gini index impurity. This selection process can automatically be done by stacking the random forest models with region/rule-encoded representations of the data. The rules are gradually combined and refined in the successive layers. Such multilayer learning architecture naturally provides the sharing and reuse of components [36]. These hierarchical combining representations based on multilayer decision trees are of more expressive power with nested CNF and DNF [29].

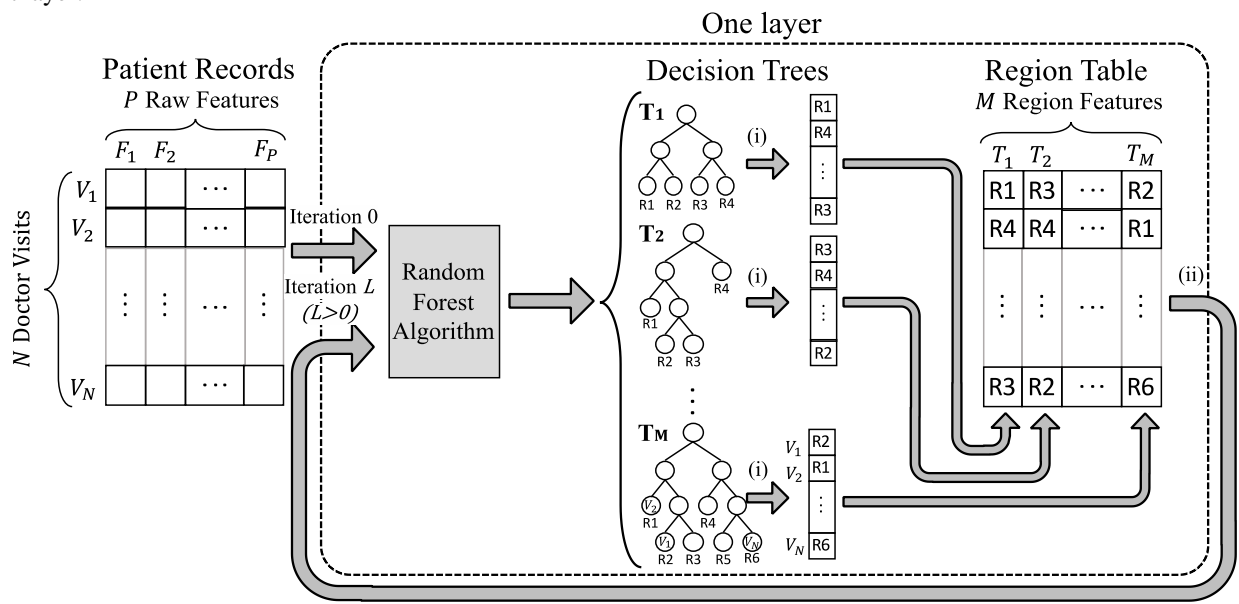

Fig. 2. Training flow of DRF with the AKI dataset

As mentioned previously, representations in DRF are in the form of DNF rules eventually. These representations can be used with other machine learning methods, such as decision tree, linear model, and Bayesian network. In Section IV, we demonstrate and explain how to fit a lasso regression model with these region representations and calculate odds ratios for various DDIs and DDXs represented as logic expressions. In short, DRF compresses original input features and then transforms them into region-wise representations layer by layer. This procedure would generate several region tables corresponding to individual layers. Hence, these region tables can be used in further tasks such as fitting a classification model or prediction.

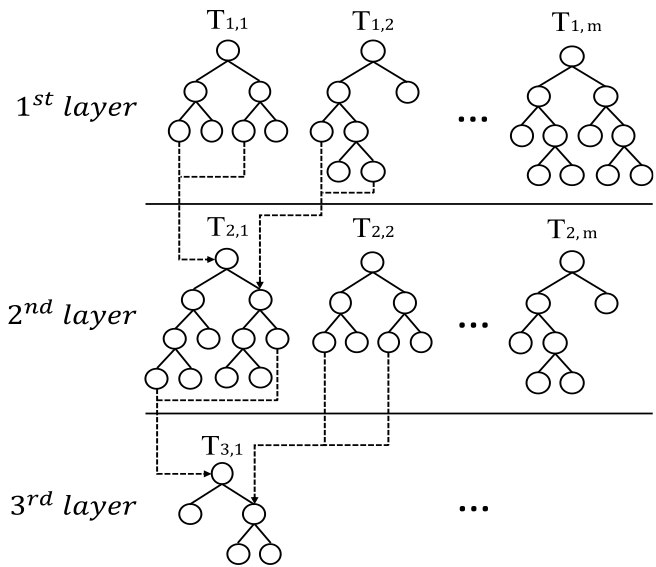

Fig. 3. Learning region-wise representations layer by layer

## IV. EXPERIMENTAL RESULTS

### A. Data Source

The data used in this study is Longitudinal Health Insurance Databases 2000, randomly sampled one million beneficiaries from original Taiwan's National Health Insurance Research Database (NHIRD) [38]. Subjects with age more than 18 years were enrolled. We excluded participants with regular dialysis (hemodialysis or peritoneal dialysis) before January 1, 2000, and those who died before April 1, 2000. We investigate subjects with at least one disease related to AKI, including diabetes mellitus, hypertension, hyperlipidemia, chronic kidney disease, heart failure, atrial fibrillation, peripheral artery disease, coronary artery disease, cerebrovascular disease (ischemic stroke or hemorrhagic stroke), chronic lung disease (chronic obstructive pulmonary disease or asthma), gout, autoimmune disease, malignancy, dementia, and psoriasis. We included 40 drug groups including histamine-2 blockers, proton pump inhibitors, anti-arrhythmic drugs, antidepressants, antiepileptics, antiplatelets, warfarin, antipsychotic agents, benzodiazepines, medications for chronic obstructive pulmonary disease treatment, digoxin, loop diuretics, thiazide, potassium-sparing diuretics (spironolactone), blood glucose-lowering drugs (insulin, acarbose, metformin, sulfonylurea, thiazolidinedione), blood pressure-lowering drugs (ACEI, ARB, beta-blocker, calcium channel blocker), hypnotics, immunosuppressive agents, lipid-lowering drugs (fibrate, statin), osteoporosis drugs, anti-parkinsonism drugs, uric lowering drugs, antibiotics (Beta-lactam, Macrolide, Quinolone, Tetracycline), antifungal drugs, anti-viral drugs, NSAID (traditional NSAID, cyclooxygenase-2 inhibitors), other pain killers (opioid), and steroid. The identified comorbidities and drug prescriptions were recorded in each clinic visit. Participants were followed up until their deaths, the occurrence of AKI events, or until December 31, 2008, whichever comes first. There are 61,663 patients/observations.

### B. Performance Evaluation

Again, the DRF model is a kind of representation learning method derived from tree-based methods and is constructed in a layer-wise fashion. Here, we compare the rule representations learned from DRF with raw features using single decision tree methods and elastic net [41]. We also consider other state-of-the-art learning algorithms with raw data, including random forest [8], XGBoost [32], and MLP [34].

The complexity of rules is determined by the number of tree nodes. We empirically set up a total of three layers in DRF, whose first and second layers were purposed to obtain more information, and the last was used to filter out excessive information which might include noise in the data. There were 100 trees in each layer. In the first two layers, the diversity of trees in one layer was formatted in the mixed numbers of leaf nodes, which were between 2 to 11. In the last layer, the diversity of trees was limited to grow into 3 leaf nodes per tree.

We used the R programming language [42] along with its packages to conduct experiments. R Package rpart, C50, RWeka built-in JRip, xgboost, ranger, keras, and H2O are used to learn CART, C4.5 (and rule-based C5.0), RIPPER, XGBoost, random forest, MLP, and elastic net models, respectively. The number of trees in the random forest, XGBoost, and each layer of the DRF was empirically set to 100. MLP was formed by two hidden layers with 16 and 8 neurons. In the experiments, due to the target variable is with imbalanced classes, we chose the area under the ROC curve (AUC) as the performance evaluation measure. Table I shows the testing AUC comparison among different learning algorithms.

The result shows that the proposed DRF with different classifiers perform better compared to other black-box models trained with raw data. Both CART and elastic nets trained with DRF's representations perform comparatively better than these two algorithms trained with the raw data. In most cases, models with DRF's data representations have more expressive power than those with raw data. It also suggests that learning DRF with multi-level rule representations with more layers may be able to identify more patterns (i.e. DDIs and DDXs) and thus result in higher accuracy of predictions.

### C. Rule Learning for Model Interpretability

Table II shows that the rules (drug-drug and drug-disease interactions) learned from rule-based learning algorithms, including CART, C4.5, rule-based C5.0, and RIPPER. According to the experiment in Section IV, the elastic net trained with DRF in the 2$^{nd}$ layer has the highest AUC. Therefore, we compare the top five crucial rules from this model in Table III. On the other hand, DRF is trained layer by layer, and thus the rules can express the medical history and medications into DNF, as shown in Table III. Rules in DNF are able to represent more complicated representations (DDIs/DDXs).

TABLE I. AUC COMPARISON AMONG DIFFERENT COMBINATIONS OF DATA REPRESENTATIONS AND ALGORITHMS

| Data | Algorithm | | | |
|---|---|---|---|---|
| | *CART [43]* | *C4.5 [26]* | *C5.0 [26]* | *RIPPER [44]* |
| Raw data | 0.723 | 0.760 | 0.760 | 0.682 |
| | *Elastic net [41], [45]* | *Random forest [46]* | *XGBoost [32]* | *MLP [47]* |
| Raw data | 0.755 | 0.769 | 0.762 | 0.782 |
| | *DRF + CART [43]* | *DRF + C4.5 [26]* | | *DRF + Elastic net [41], [45]* |
| DRF in 1st layer | 0.726 | 0.750 | | 0.779 |
| 2nd layer | 0.718 | 0.734 | | 0.782 |
| 3rd layer | 0.760 | 0.757 | | 0.766 |

TABLE II. RULES COMPARISON OF RULE-BASED MODELS

| Algorithm | Rule | # of records (Positive) |
|---|---|---|
| CART | CCB (Calcium channel blockers) & DIU (Loop diuretics) | 6,010(177) |
| C4.5 | Age = 60-79 & H2B (H2 blockers) & HTN (Hypertension) & Economic level in (2nd, 3rd) & COP (COPD medications) | 34(9) |
| Rule-based C5.0 | MET & Age = 18-39 | 6,433(33) |
| RIPPER | DM (Diabetes mellitus) | 120(117) |

TABLE III. RULES COMPARISON OF ELASTIC NET TRAINED FROM DRF IN THE 2ND LAYER

| Importance | Rule | # of records (Positive) |
|---|---|---|
| 1st important (L2.T3=r1) | MET | NSA | (NSA & P_CKD) | BZD | (BZD & PSD) | ACE | (ACE & DIU) | 27,307(377) |
| 2nd important (L2.T2=r2) | (SUL & P_CKD & STA) | SUL | (SUL & P_CVAD) | PAI | STA | |(STA & P_CVAD) | (SUL & P_CKD) | 87,154 (1,891) |
| 3rd important (L2.T2=r1) | P_CKD | 572(13) |
| 4th important (L2.T1=r2) | PAI & P_CKD | 41,835 (1,444) |
| 5th important (L2.T1=r1) | ACE | MET | PAI | 45,891(460) |

*D. AKI Risk in DDI or DDX*

The experimental result also suggests that each rule can cover a sub-group of patient records, and subjects with these rules are at high risk of AKI. Fig. 4 shows that the top five groups/rules/regions that denote those who may have a high risk of AKI. The numbers below black points are estimates of odds ratios of a logit model. Fig. 4 suggests that, for example, those who have taken MET (Metformin), NSA (NSAIDs), NSA with CKD (chronic kidney disease), BZD (Benzodiazepines), a combination of BZD and PSD (Spironolactone), ACE (ACEI), or combination of ACE (ACEI) and DIU (loop diuretics) are 3.19 times more likely to have AKI than the others. According to the rules from the raw data, we demonstrate traditional NSAIDs as one of the top important factors related to AKI. Reported from systematic review and meta-analysis can confirm our finding of AKI risks in traditional NSAID users [48]. Besides, metformin (MET), which is the most common oral antidiabetic agents for type 2 diabetes treatment, is also associated with AKI episode. Metformin associated lactic acidosis is usually reported in patients with AKI [49]. However, the metformin and AKI association should be interpreted as a consequence, rather than causality. The concurrent use of ACEI and loop diuretics may potentially increase the risk of AKI, especially in hypovolemic status or additional NSAID exposure [50]. Our finding of increased AKI risk in concomitant use of benzodiazepines and spironolactone is novel. The AKI episode after benzodiazepines used was demonstrated in the case report [51]. Further study to evaluate the risk of AKI in co-current use of benzodiazepines and spironolactone is needed.

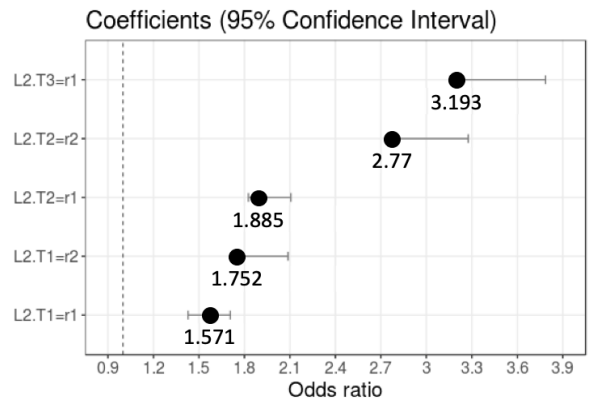

Fig. 4. Odds ratio with 95% confidence intervals generated from the elastic net model, which was trained with the 2nd layer of DRF

V. CONCLUSION

In this paper, we propose a novel rule-based feature learning technique, Deep Rule Forests, that helps discover DDIs and DDXs that may potentially induce AKI. We found that our proposed approach performs comparatively better than typical rule-based algorithms and the update-to-date black-box algorithms in terms of model interpretability and accuracy of prediction, respectively. This work could help researchers and physicians to identify adverse drug reactions when drugs are used alone or in combinations. Also, we can obtain multiple sets of DDIs and DDXs that may cause AKI as well as exhaustive rules if there are multiple layers in the DRF model.